\documentstyle[preprint,prl,aps]{revtex}
\begin{document}
\draft
\title{Spin-Excitation-Instability-Induced Quantum Phase Transitions 
	in Double-Layer Quantum Hall Systems }
\author{Lian Zheng, R.J. Radtke, and S. Das Sarma} 
\address{Department of Physics,
University of Maryland, College Park, Maryland 20742-4111}
\date{\today}
\maketitle
\begin{abstract}
We study intersubband
spin density collective modes 
in double-layer 
quantum Hall systems at $\nu=2$ within the time-dependent 
Hartree-Fock approximation. We find that these intersubband 
spin density excitations may soften
under experimentally accessible conditions, signaling a
phase transition
to a new quantum Hall state
with interlayer inplane antiferromagnetic spin correlations. 
We show that this novel canted
antiferromagnetic phase is energetically stable and that the phase transition
is continuous. 

\end{abstract}
\narrowtext
Electron systems in confined geometries exhibit a richer 
variety of physical properties than their higher-dimensional counterparts
due to enhanced interaction 
effects in reduced dimensions. Interaction in a low-dimensional 
system
does not merely result in stronger
renormalization of physical quantities, but can in many cases 
drive the system into completely new 
phases with peculiar properties. 
For a two-dimensional  
electron gas in a perpendicular
magnetic field, the interaction effects are especially important
because of Landau level quantization. 
When electrons are entirely restricted 
to the lowest Landau level by a large magnetic field, 
electron-electron interaction completely dominates
the properties of the system as the electron
kinetic energy is
quenched to an unimportant constant.
One of the most interesting phenomena in this strongly-correlated system
is the quantum Hall effect (QHE), which has attracted 
a great deal of experimental and theoretical interest 
during the last fifteen years \cite{smg1}. 
Recent advances in materials growth techniques have 
made it possible to fabricate high-quality double-layer
two-dimensional electron systems with the electrons confined 
to two parallel planes separated by a distance comparable to that
between electrons within a plane.
With the introduction of this
layer degree of freedom, many qualitatively new 
effects due entirely to interlayer correlations appear 
\cite{kun1,nuh,pit1}.
In this letter, we present a theoretical study of
the intersubband spin-density-wave (SDW) excitations and 
the associated phase transitions  
in double-layer electron systems at a total Landau level filling
factor $\nu=2$.  
The intersubband SDW dispersion is evaluated in
the time-dependent Hartree-Fock approximation 
\cite{kal1}.
We find that 
the intersubband SDW modes could soften 
under experimentally accessible conditions, leading
to a phase transition to a novel QHE state 
with interlayer inplane antiferromagnetic 
spin correlations $\langle S^x_L\rangle=-\langle S^x_R\rangle\neq0$ 
($S$ is the electron spin operator, $\hat x$ is a direction
parallel to the 2D plane
with the magnetic field along the $\hat z$ 
direction, and $L$ and $R$ denote the 
left and right layers,
respectively).
Using a mean-field approximation,
we are able to show that this antiferromagnetic phase is
energetically stable and that the phase transition is continuous.
We are, therefore, predicting a new quantum phase transition
to a novel canted antiferromagnetic state 
in the double-layer system which occurs at zero temperature
as system parameters (such as interlayer separation) are varied.
Our findings seem to be consistent with recent 
inelastic light scattering measurements,\cite{ap1}
where a remarkable softening of the long wavelength SDW mode 
in a $\nu=2$ double quantum well system has been observed.

There has been a lot of work on double-layer QHE systems.
Most studies \cite{kun1,nu1}, however, have focused on
$\nu=1$ (with some work\cite{nuh} on $\nu=1/2$),
leaving the $\nu=2$ state essentially uninvestigated. 
Our work shows that the $\nu=2$ QHE state,
where the spin and the layer index compete with each other,
has non-trivial magnetic properties.
Although $\nu=1$ and 
$\nu=2$ double-layer 
QHE states exhibit some similarities such as the softening
of the low energy collective excitations under certain conditions,
there are important differences between them.
At $\nu=1$, the spin degree of freedom is normally frozen out
because of the SU(2) symmetry of the Coulomb interaction.
The relevant low energy excitations in the $\nu=1$ QHE state are
therefore intersubband charge-density-wave excitations,
and the properties of the system are determined by the interplay between
the interlayer tunneling energy and the Coulomb interaction energy.
At $\nu=2$,
both the spin degree of 
freedom and the layer degree of freedom are relevant,
and the low energy excitations are intersubband SDW excitations.
Consequently, the properties of the system are determined 
by the interplay among the tunneling energy, the Zeeman energy, 
and the Coulomb interaction energy.
At $\nu=1$, the mode softening destroys the QHE \cite{nu1}
because beyond the critical layer separation the system is effectively 
a pair of isolated layers with compressible half-filled Landau level
states,
while at $\nu=2$, the QHE prevails in all phases 
due to the existence of 
incompressible filled Landau level states
with charge excitation gaps, even at $d\rightarrow\infty$.
The spin mode softening and the associated quantum 
phase transition to the canted antiferromagnetic state
at $\nu=2$ are, however, experimentally 
observable through inelastic
light scattering experiments \cite{ap1}.

In this work, we employ two different approaches to study
the spin excitation instabilities 
in double-layer systems at $\nu=2$.
These two approaches provide complementary information: 
one approach \cite{kal1} deals with the collective excitations while
the other \cite{wig1} deals with the ground state properties.
Both approaches are based on the Hartree-Fock 
approximation.  
In single-layer integer QHE systems, 
calculations \cite{kal1} in the Hartree-Fock approximation  
agree well with experiments \cite{ap2}.
In double-layer systems, the Hartree-Fock approximation
is less accurate because the Coulomb interaction potentials are 
more complicated.  Nevertheless, we expect 
the Hartree-Fock approximation 
to remain reasonably good for a double-layer system at $\nu=2$, 
since the Hartree-Fock ground state, 
which is non-degenerate and separated in energy from higher levels,
is a good approximation for the real many-body ground state
at $\nu=2$ due to the existence of filled Landau levels and
charge excitation gaps.
Because our calculations employ 
realistic Coulomb interaction potentials (including finite
well-thickness corrections \cite{rk1})
and incorporate  
interlayer tunneling and Zeeman splitting,
we expect our results to be not only qualitatively correct
but also quantitatively reliable. 

The Hamiltonian of the system
is ${\cal H} = {\cal H}_0 + {\cal H}_{\rm I}$ with
\begin{equation}
{\cal H}_0=-{\Delta_{\rm sas}\over2}\sum_{\alpha\sigma}\ \left(
C^\dagger_{1\alpha\sigma}
C_{2\alpha\sigma}+h.c.\right)-{H\over2}\sum_{i\alpha\sigma}{\sigma}
C^\dagger_{i\alpha\sigma}C_{i\alpha\sigma},
\end{equation}
where $C_{i\alpha\sigma}$ annihilates an electron in the lowest Landau
level
in layer $i$ ($i=1,2$) with spin $\sigma$ ($\sigma = \pm 1$)
in the direction of the perpendicular field and
with intraLandau level index $\alpha$.
Interlayer tunneling induces the symmetric-antisymmetric energy
separation
$\Delta_{\rm sas}$.
The Coulomb interaction part of ${\cal H}$ is
\begin{eqnarray}
{\cal H}_{\rm I}=&&{1\over2}\sum_{\sigma_1\sigma_2}\sum_{ij}
\sum_{\alpha_1\alpha_2}{1\over\Omega}\sum_{\bf q}
V_{ij}(q)e^{-q^2l_o^2/2}e^{iq_x(\alpha_1-\alpha_2)l_o^2}
\nonumber \\
&&\times C^\dagger_{i\alpha_1+q_y\sigma_1}C^\dagger_{j\alpha_2\sigma_2}
C_{j\alpha_2+q_y\sigma_2}C_{i\alpha_1\sigma_1},
\label{equ:hi}
\end{eqnarray}
where $\Omega$ is the area of the sample.
The interaction potentials are
$V_{ij}={2\pi e^2/\epsilon q}F_a(q)$ for $i=j$ and
$V_{ij}={(2\pi e^2/\epsilon q})e^{-qd}F_e(q)$ for $i\ne j$.
The finite-layer-thickness form factors $F_{a(e)}$   
used in our calculations
are taken from ref. \cite{wig1}.

We use $|\alpha\mu\sigma\rangle$ to denote the eigenstates of ${\cal H}_0$,
where $\mu=0,1$ labels the symmetric and antisymmetric subbands.
There are two intersubband
SDW excitations which correspond to transitions 
$|0\uparrow\rangle\leftrightarrow|1\downarrow\rangle$ and 
$|0\downarrow\rangle\leftrightarrow|1\uparrow\rangle$.
In the absence of interaction, these  
modes have excitation energies 
$|\Delta_{sas}\pm\Delta_z|$.
The interaction renormalizes the excitation energies 
in two ways. One is due to the loss of exchange energy when
an electron is excited to a higher but empty level, which raises 
the excitation energies. The other is an excitonic attraction 
between the electron excited to the higher level 
and the hole it leaves behind, which lowers the excitation 
energies.
In diagrammatic perturbation theories, 
the effect of the exchange energy 
on the excitation energies is accounted for by
including the 
corresponding self-energy in the electron Green's functions,
and the effect of the 
excitonic attraction is represented by vertex corrections. 
The direct Hartree term does not influence the SDW excitations
because the Coulomb interaction is spin-rotationally invariant. 
Since the Coulomb interaction potentials are subband-index dependent,
they introduce mode-coupling between the two branches of the intersubband
SDW excitations. This mode-coupling pushes down the frequency
of the low-lying excitation
and helps mode softening.
The intersubband SDW excitation spectra
are obtained as the poles of the retarded spin-density response
function \cite{kal1}
\begin{equation}
\chi(q,\omega)=-i\int_0^\infty e^{i\omega t}
\langle[\rho_{SD}({\bf q},t),\rho_{SD}^\dagger(-{\bf q},0)]
\rangle,
\label{equ:x1}
\end{equation}
where the intersubband SDW operator $\rho_{SD}$ is defined as follows.
In the ferromagnetic ground state, {\it i.e.}
when $|0\uparrow\rangle$ and $|1\uparrow\rangle$ are occupied, 
$\rho_{SD}({\bf r})=\sum_{\mu}\psi_{\mu\uparrow}^\dagger({\bf r})
\psi_{1-\mu\downarrow}({\bf r})$, 
where $\psi_{\mu\sigma}$  annihilates
an electron in subband $\mu$ with spin $\sigma$.
In the symmetric ground state, {\it i.e.}
when $|0\uparrow\rangle$ and $|0\downarrow\rangle$ are occupied,
$\rho_{SD}({\bf r})=\sum_{\sigma}\psi_{0\sigma}^\dagger({\bf r})
\psi_{1-\sigma}({\bf r})$.
Eq. (\ref{equ:x1}) is evaluated 
in the time-dependent Hartree-Fock approximation 
\cite{kal1},
which we adapt to double-layer systems
and, for simplicity, ignore all the higher 
Landau levels. As argued above, this should be a 
good approximation for our problem.

In Fig. \ref{fig1}, we show the dispersion of the intersubband SDW 
in the ferromagnetic ground state.
As mentioned earlier, there are two intersubband SDW modes 
$\omega_{\pm}(q)$ corresponding 
respectively to transitions $|0\uparrow\rangle\rightarrow
|1\downarrow\rangle$ and $|1\uparrow\rangle\rightarrow
|0\downarrow\rangle$.
The frequencies $\omega_{\pm}$ increase as functions of $q$,
approaching asymptotic values $\omega_{\pm}(q\rightarrow\infty)
=\omega_{\pm}^0+|v_x|$,
where $\omega_{\pm}^0$
are the non-interacting excitation energies  
and $v_x$ is the exchange energy of an electron
in the ground state.
Mode coupling, which 
pushes down $\omega_-(q)$ and hence
helps mode softening,
is most visible at $q\rightarrow0$.
At zero layer separation, mode-coupling disappears, 
and we recover previously known results \cite{kal1,nu1}.
The dispersions in Fig. \ref{fig1} are shown for two  
different input parameters. In one case, there
is no mode softening 
($\omega_{\pm}(q)>0$), and in the other, there is mode softening
($\omega_-(q\rightarrow0)<0$).
The mode softening signals that the 
ground state is unstable against 
spontaneous generation of intersubband
SDW excitations.
In Fig. \ref{fig2}, we show the intersubband SDW dispersion
in the symmetric ground state. The results are qualitatively
similar to
those in Fig. \ref{fig1}. 
The important thing to notice is that
there is mode softening here as well.
We emphasize that our calculated SDW
dispersion is experimentally measurable through 
depolarized inelastic light scattering experiments.
In fact, the softening of the intersubband SDW excitation
in a double-layer system at $\nu=2$ (but NOT at
$\nu=4$) may have already been observed in
a recent experiment \cite{ap1}. 

The mode softening at
$q\rightarrow0$ suggests
that the new phase has a broken symmetry:
$\langle\rho_{SD}(q=0)\rangle\ne0$,
where $\rho_{SD}(q)$ is the intersubband SDW operator defined earlier.
Since $\langle\rho_{SD}(q=0)\rangle
=N_\phi
[\langle S_L^x\rangle-
\langle S_R^x\rangle$], where $N_\phi$ is the Landau level degeneracy,
the new phase is in fact characterized by
an interlayer canted antiferromagnetic spin correlation. 
The phase diagram can be obtained by tracing the points where the
mode softening occurs.
The results are shown in Fig. \ref{fig3}. 
There are three phases
in a double-layer QHE system at $\nu=2$:
a ferromagnetic phase 
where $|0\uparrow\rangle$ and $|1\uparrow\rangle$ are occupied,
a symmetric phase where $|0\uparrow\rangle$ and $|0\downarrow\rangle$ 
are occupied (it may also be viewed as ferromagnetic in
the sub-space associated with the layer degrees of freedom \cite{kun1}),
and the antiferromagnetic phase.
The symmetric phase exists for $\Delta_{sas}>\Delta_z$
and $d<d_{c1}$, the antiferromagnetic phase
exists for $\Delta_{sas}>\Delta_z$
and $d_{c1}<d<d_{c2}$,
and the ferromagnetic phase exists for either $\Delta_z>\Delta_{sas}$
or $d>d_{c2}$. 
The ferromagnetic phase is favored when  
$\Delta_z$ is increased, while
the symmetric phase is favored when 
$\Delta_{sas}$ is increased.
It should be noted that, for a given $\Delta_{sas}$,
$d_{c1}$ is considerably smaller 
than the critical layer separation 
where the charge density excitation in the $\nu=1$ state
becomes soft \cite{nu1}.
The reason for this is the absence of the Hartree contribution to
SDW excitations.
An intuitive explanation on the existence of the canted 
antiferromagnetic phase
is that in this  phase the system can take advantage of the
tunneling energy ( the kinetic energy in the perpendicular direction)
by having an inter-layer antiferromagnetic spin correlation --- an example  
of the
super-exchange induced 
antiferromagnetic correlation. The reason
the antiferromagnetic ordering is canted is to minimize Zeeman energy.
We emphasize that our predicted phase diagram should be
experimentally measurable.

The antiferromagnetic ground state
can also be calculated within a mean-field approximation.
The approach we employ
is the same as that used
to study magnetic-field-induced Wigner solid phase in 2D systems 
\cite{wig1}, except that
we only look for uniform solutions and allow 
for the possibility of a
non-zero antiferromagnetic order parameter 
($\langle\rho_{SD}(q=0)\rangle\ne0$).
The energies obtained from this mean-field theory show
that the antiferromagnetic phase 
is energetically stable.  
In Fig. \ref{fig4}, we show
the antiferromagnetic order parameter obtained 
in this mean-field approximation.
Several features are obvious from this figure:
(i) The antiferromagnetic phase exists only at intermediate
layer separations. (ii) The range of layer separations in which 
the antiferromagnetic phase exists shrinks with increasing Zeeman energy.
(iii) The phase transitions are continuous.
This mean-field theory also provides a way to derive the phase diagram,
and the phase diagram obtained in this manner 
is identical to the one obtained from the softening of 
the intersubband SDW excitations, providing a confirmation
of our predictions.

Our calculations at $\nu=2$ are well controlled in the absence 
of Landau level mixing.
We expect that the Landau level mixing is negligible
at $\nu=2$, but may not be negligible at $\nu=6$
(otherwise, the results presented here would be qualitatively
valid at $\nu=6$ \cite{rk2}).
We would like to emphasize that the Hartree-Fock approximation 
fails completely at non-integer filling factors,
because the ground states are macroscopically 
degenerate. Nevertheless, some qualitative speculations 
for non-integer filling factors can be made.
For example, when $\nu$ is changed away from $2$, 
the presence of screening, which reduces the electron-hole 
excitonic attraction,
may increase the intersubband SDW energy and prevent the mode softening. 
Thus, the antiferromagnetic phase would be unstable 
away from $\nu=2$.
This is precisely the experimental observation in ref.\cite{ap1}
where a sharp minimum in the SDW energy is found at $\nu=2$.

In summary, 
we have studied the instabilities induced 
by the softening of intersubband
spin-density-wave excitations 
in double-layer 
quantum Hall systems at $\nu=2$ within the time-dependent 
Hartree-Fock approximation. The intersubband 
spin-density excitation modes soften
under experimentally accessible conditions and lead to
a novel quantum Hall state
with interlayer planar antiferromagnetic spin correlations. 
We show, in a mean-field approximation, that this
planar antiferromagnetic phase is energetically stable and 
that the phase transition
is continuous. 
We, therefore, predict the existence of a novel 
canted antiferromagnetic phase under suitable conditions
in between the symmetric and the ferromagnetic phases in
a double-layer QHE sample at $\nu=2$.

The authors 
thank Dr. A. Pinczuk 
for helpful discussions on the experimental data. 
This work is supported by the
U.S.-ARO and the U.S.-ONR.

\begin{figure}
\caption{ 
Inter-subband SDW dispersion $\omega_{\pm}(q)$
in the ferromagnetic (FM) phase at $\nu=2$.
The solid lines are for $\Delta_{sas}=0.02 e^2/\epsilon l_o$,
and the dashed lines are for $\Delta_{sas}=0.1 e^2/\epsilon l_o$.
Other parameters are the same for both the solid and dashed lines:
the Zeeman energy $\Delta_z=0.01e^2/\epsilon l_o$, 
the layer separation $d=1.15l_o,$
and the well-thickness $d_w=0.8l_o$.
Notice that $\omega_-(q\rightarrow0)<0$ for $\Delta_{sas}
=0.1 e^2/\epsilon l_o$
(dashed line).
}
\label{fig1}
\end{figure}

\begin{figure}
\caption{ 
Inter-subband SDW dispersion $\omega_{\pm}(q)$
in the symmetric (SYM) phase at $\nu=2$.
The solid lines are for layer separation $d=0.85l_o$ and the dashed lines
are for $d=0.95l_o$.
Other parameters are the same for both the solid lines and dashed lines:
$\Delta_z=0.08e^2/\epsilon l_o,\  \Delta_{sas}=0.35e^2/\epsilon l_o$,
$d_w=0.8l_o$.
Notice that $\omega_-(q\rightarrow0)<0$ for $d=0.95l_o$
(dashed line).
}
\label{fig2}
\end{figure}

\begin{figure}
\caption{ 
Phase diagrams at $\nu=2$ for two different Zeeman energies:
(a) $\Delta_z=0.01e^2/\epsilon l_o$ and
(b) $\Delta_z=0.08e^2/\epsilon l_o$.
The well-thickness $d_w=0.8l_o$ in both cases.
Three phases are present: a symmetric phase (SYM),
a ferromagnetic phase (FM), and an antiferromagnetic phase (AF).
The `$\times$' in (a) denotes the experimental sample parameters
of Ref. 6 (with a magnetic field $B=1.3T$),
where the measured SDW energy has a sharp minimum
at $\nu=2$ with a value of $0.04$ meV
which is comparable to the experimental temperature of $0.6K$.}
\label{fig3}
\end{figure}

\begin{figure}
\caption{ 
Antiferromagnetic order parameter $\langle\rho_{SD}(q=0)\rangle/N_\phi$
versus layer separation $d$ for the indicated Zeeman energies,
where $N_\phi$ is the Landau level degeneracy and $\rho_{SD}(q)$
is the intersubband spin-density operator defined
in the text. The well-thickness $d=0.8l_o$.
}
\label{fig4}

\end{figure}


\begin{references}
\bibitem{rk1}
The finite well-thickness corrections, as well as other detailed
corrections to interaction potentials, have quantitative effects only.
They do not change any results qualitatively.
\bibitem{rk2}
On the other hand, the situation at $\nu=4$ is very different
from the situation at $\nu=2$,
since the inter-Landau level excitation energies are
comparable to the cyclotron energy.
\end{references}
\end{document}